\documentclass[11pt,a4paper]{article}

\def\AmSTeX{\leavevmode\hbox{$\mathcal A\kern-.2em\lower.376ex%
        \hbox{$\mathcal M$}\kern-.2em\mathcal S$-\TeX}}
\usepackage{graphicx}
\usepackage{amsmath}        
\usepackage{amssymb}
\usepackage{mathrsfs}       
\usepackage{bm}             
\usepackage[amsmath,thmmarks,hyperref]{ntheorem}
\newif\ifpdf \pdftrue
\ifx\pdfoutput\undefined \pdffalse \fi \ifx\pdfoutput\relax
\pdffalse \fi
\ifx\texonly\undefined\let\texonly\relax\fi
\ifx\endtexonly\undefined\let\endtexonly\relax\fi \texonly
  \let\htmlonly\iffalse
  \let\endhtmlonly\fi
\endtexonly

\htmlonly
  
\endhtmlonly
\texonly
  \usepackage{makeidx}

  \usepackage{multicol}
  

\endtexonly
\title{}
\author{\thanks{}}
\date{}
\begin{document}

\title{Semi-leptonic and Non-leptonic $B$ meson decays to charmed mesons}

\author{Fu Hui-feng(付慧峰), Wang Guo-Li(王国利)\footnote{gl\_wang@hit.edu.cn}, Wang Zhi-Hui(王志会), Chen Xiang-Jun(陈相君) \\
{\it \small   Department of Physics, Harbin Institute of Technology,
Harbin, 150001, China} }


\maketitle

\baselineskip=20pt
\begin{abstract}
\noindent We study the semi-leptonic and non-leptonic $B$ weak
decays which are governed by the $B\rightarrow D^{(*)}$
transitions. The branching ratios, CP asymmetries (CPA) and
polarization fractions (FA) of non-leptonic decays are investigated in
the factorization approximation. The $B\rightarrow D^{(*)}$
form factors are estimated in the Salpeter method. Our estimation on
branching ratios generally agree with the existent experimental
data. For CPA and polarizations, comparisons among the FA results,
the perturbative QCD predictions and experimental data are
made.
\end{abstract}

Recent years, more and more measurements on the physics of
charmed-meson $B$ decays have been made~\cite{Belle2,BaBar}. The
factorization approximation~\cite{Bauer} has been justified to
work well in $B\rightarrow D^{(*)}X$ decays. Till now, several
works on estimating $B\rightarrow D^{(*)}$ form factors have been
done including constituent quark model (CQM)~\cite{CQM3}, QCD sum
rules (QCDSR)~\cite{Azizi} and heavy quark symmetry
(HQS)~\cite{DBGN,NS,CGW}. In this work, we study $B\rightarrow
D^{(*)}$ precesses with the Salpeter method (instantaneous
Bethe-Salpeter method~\cite{Salpeter2}).

FA has not been justified if it still works in the CP asymmetry.
Besides, the non-factorizable contributions to CPA, calculated by
some approaches beyond the FA, such as QCD factorization
(QCDF)~\cite{QCDF1} and perturbative QCD
(pQCD)~\cite{PQCD2,PQCD3}, is still unclear. Actually the QCDF and
the pQCD predictions on CPA are quite different even have opposite
sign~\cite{LiHN}. So, it seems important to make a comparison
between the theoretical predictions and experimental data on CPA.
The polarization of the decays to two (axial) vectors is another
important observable which is also covered in this letter.

The wave functions used here which have quantum numbers $J^P=0^-$
(for $B^{0(\pm)}$ and $D^{0(\pm)}$) and $1^-$ (for $D^{*0(\pm)}$)
 are written as~\cite{Wang}
$$\begin{array}{rl}
\varphi_{0^{-}}(q_{_{P_\perp}})=&M\left[\frac{\not\!P}{M}a_1(q_{_{P_\perp}})+{a}_{2}(q_{_{P_\perp}})
+\frac{\not\!q_{_{P_\perp}}}{M}{a}_{3}(q_{_{P_\perp}})+\frac{\not\!{P}\not\!q_{_{P_\perp}}}{M^2}{a}_{4}(q_{_{P_\perp}})\right]{\gamma}_5,\notag
\\
{\varphi}^{\lambda}_{1^{-}}(q_{_{P_\perp}})=&(q_{_{P_\perp}}\cdot\epsilon^{\lambda})\left[b_1(q_{_{P_\perp}})+\frac{\not\!{P}}{M}b_2({q_{_{P_\perp}}})
+\frac{\not\!q_{_{P_\perp}}}{M}b_3({q_{_{P_\perp}}})+\frac{\not\!{P}\not\!{q_{_{P_\perp}}}}{M^2}b_4({q_{_{P_\perp}}})\right]+M\not\!{\epsilon}^{\lambda}b_5({q_{_{P_\perp}}})\notag\\
&+\not\!{\epsilon}^{\lambda}\not\!{P}b_6(q_{_{P_\perp}})
+(\not\!{q_{_{P_\perp}}}\not\!{\epsilon}^{\lambda}-q_{_{P_\perp}}\cdot{\epsilon}^{\lambda})b_7({q_{_{P_\perp}}})
+\frac{1}{M}(\not\!{P}\not\!{\epsilon}^{\lambda}\not\!{q_{_{P_\perp}}}-\not\!{P}q_{_{P_\perp}}\cdot{\epsilon}^{\lambda})b_8({q_{_{P_\perp}}}),\notag
\end{array}$$
where $a_i(q_{_{P_\perp}})$ and $b_i(q_{_{P_\perp}})$ are
functions of $q_{_{P_\perp}}^2$; $M$ is the mass of the
corresponding meson; $\epsilon^\lambda$ is the polarization
vector. $q_{_P}=\frac{q\cdot P}{\sqrt{P^2}}$ and
$q_{_{P_\perp}}=q-\frac{q\cdot P}{P^2}P$. We chose the parameters
: $m_u=0.305$ GeV, $m_d=0.311$ GeV, $m_c=1.62$ GeV and $m_b=4.96$
GeV \cite{Wang} to solve the Salpeter equation and obtained the
numerical values of wave functions.

\begin{figure}[t]
\centering
\includegraphics[width = 0.8\textwidth]{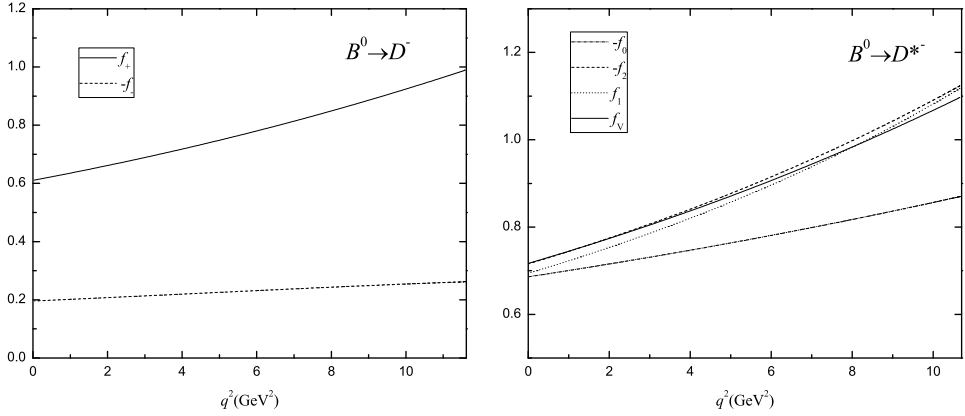}
\caption{Form factors of $B^0\rightarrow D^{(*)-}$ weak transition.}
\label{fig4-1}
\end{figure}

According to Mandelstam formalism~\cite{Mandelstam}, under the
instantaneous approximation, a transition matrix element induced
by a current $\Gamma^\mu=\gamma^\mu(1-\gamma_5)$ can be written as
\begin{equation}\label{eq1}
\langle f(P_f) |\bar{q}_1\Gamma^\mu
q_2|i(P_i)\rangle=\int\frac{d^3q_{i_{P_{i\perp}}}}{(2\pi)^3}\mathrm{Tr}\Big[\bar{\varphi}_{f}^{++}(q_{f_{P_{f\perp}}})
\frac{\not\!P_f}{M_f}L_s\Gamma^\mu\varphi_{i}^{++}(q_{i_{P_{i\perp}}})\frac{\not\!P_f}{M_f}\Big],
\end{equation}
where
$\varphi^{++}_{i(f)}\equiv\Lambda_{1}^{+}\frac{\not\!P_{i(f)}}{M_{i(f)}}\varphi\frac{\not\!P_{i(f)}}{M_{i(f)}}\Lambda_{2}^{+}$
and $\bar{\varphi}=\gamma_0\varphi^{\dag}\gamma_0$. $
\Lambda_{j}^{\pm}(p_{_{P_\perp}})\equiv
\frac{1}{2\omega_{j}}[\frac{\not\!{P}}{M}\omega_{j}\pm(\not\!p_{_{P_\perp}}+(-1)^{j+1}m_{j})]
$ with $\omega_{j}=\sqrt{m_{j}^{2}-p_{_{P_\perp}}^{2}}$, where $j=1$
for quark and $j=2$ for anti-quark.
$L_s=\frac{M_f-\omega_{1f}-\omega_{2f}}{P_{f_{P_i}}-\tilde{\omega}_{1}-\tilde{\omega}_{2}}\Lambda_{1}^{+}(p_{1f_{P_{i\perp}}})$;
$q_{f_{P_{f\perp}}}=q_{f_{P_{i\perp}}}-\frac{q_{fP_{i\perp}}\cdot
P_{fP_{i\perp}}}{M_f^2} P_f+ s
(\frac{1}{M_i}P_i-\frac{P_{fP_{i}}}{M^2_f}P_f)$, with
$s=\frac{m_{2f}}{m_{1f}+m_{2f}} P_{fP_{i}}-\omega_{2i}$. The
transition matrix element can be expressed with form factors by the
following decompositions:
\begin{equation}\label{eq2}\begin{array}{rl}\langle
D|\bar{q}_1\Gamma^\mu
q_2|B\rangle\equiv& f_+(Q^2)P^\mu+f_-(Q^2)Q^\mu, \\
\langle D^{*} |\bar{q}_1\Gamma^\mu q_2|B\rangle\equiv&
i\frac{2}{M_i+M_f}f_V(Q^2)\varepsilon^{\mu\epsilon^* P_i
P_f}+f_1(Q^2)\frac{\epsilon^*\cdot P_i}{M_i+M_f} P^\mu \\&
+f_2(Q^2)\frac{\epsilon^*\cdot P_i}{M_i+M_f} Q^\mu
+f_0(Q^2)(M_i+M_f)\epsilon^{*\mu},
\end{array}\end{equation}
where $P\equiv P_i+P_f$ and $Q\equiv P_i-P_f$. $f_\pm(Q^2)$,
$f_V(Q^2)$ and $f_i(Q^2)\  (i=0,1,2)$ are the form factors, we
draw them in Fig.~\ref{fig4-1}. The form factors of
$B^{-}\rightarrow D^{(*)0}$ are almost the same as them.

Now we turn to study the $B\rightarrow D^{(*)}l\nu_l$ decays and
$\bar{B}^0\rightarrow D^{(*)+}+L^-$ decays, where $L$ denotes a
light meson, and $\bar{B}^0(B^-)\rightarrow
D^{(*)+}(D^{(*)0})+D_{d(s)}^-$ decays, in the framework of FA.
Starting from the effective Hamiltonian of interested processes,
the decay amplitudes are gathered as
follows\cite{Buras}--\cite{Du1}:
\begin{equation}\label{eq3}\begin{array}{rll}
\mathcal{M}=&\frac{G_F}{\sqrt{2}}V^*_{cb}\bar{\nu}_l\gamma^\mu(1-\gamma_5)l\langle
D^{(*)} |\bar{b}\gamma_\mu(1-\gamma_5)c|B\rangle,
&\mathrm{for}~~B\rightarrow D^{(*)}l\nu_l,\\
\mathcal{M}=&\frac{G_F}{\sqrt{2}} V_{cb}V^*_{uq} a_1 A,
&\mathrm{for}~~ \bar{B}^0\rightarrow D^{(*)+}L^-,\\
\mathcal{M}=&\frac{G_F}{\sqrt{2}}\Big\{\lambda_c
a_1+\sum\limits_{p=u,c}\lambda_p\big[a^p_4+a^p_{10}+\xi(a^p_6+a^p_8)\big]\Big\}A,
&\mathrm{for}~~B\rightarrow
D^{(*)}D_{d(s)},
\end{array}\end{equation}
where $A=\langle D^{(*)} |\bar{c}\Gamma^\mu b|B\rangle\langle X
|\bar{q}\Gamma^\mu p|0\rangle$, $\lambda_p\equiv V_{pb}V^*_{pq}$ and
$V_{pq}$ is the CKM matrix element with $p=u,c$ and $q=d,s$. $a_i$ are the combinations of Wilson coefficients
$C_i(\mu)$: $a_{2i-1}\equiv C_{2i-1}+\frac{C_{2i}}{N_c},\ \
a_{2i}\equiv C_{2i}+\frac{C_{2i-1}}{N_c}$ with $N_c=3$. In the
amplitudes, $a_1$ term corresponds to the contribution from the
color-favored tree diagram, $a_{4,6}^p$ ($a_{8,10}^p$) correspond to the QCD (electroweak)
penguin contributions where $a_i^p\equiv a_i+I_i^p$ with
$I_4^p=I_6^p=\frac{\alpha_s}{9\pi}\big\{C_1[\frac{10}{9}-G(m_p,k^2)]\big\}$,
$ I_8^p=I_{10}^p=\frac{\alpha_e}{9\pi
N_c}\big\{(C_1+C_2N_c)[\frac{10}{9}-G(m_p,k^2)]\big\}$ and
$G(m_p,k^2)=-4\int_0^1x(1-x)\mathrm{ln}\frac{m_{p}^2-k^2x(1-x)}{m_b^2}dx$,
 where $k$ is the penguin momentum transfer. We
take it to be
$k^2=\frac{m_b^2}{2}(1+(m_{\bar{q}_x}^2-m_{q}^2)(1-\frac{m_{\bar{q}_x}^2}{m_b^2})/M_X^2
+(m_{q}^2+2m_{\bar{q}_x}^2-M_X^2)/m_b^2)$ as did in Ref.~\cite{Du1}.
The $\xi$ in Eq.~(\ref{eq3}) arises from the right-handed currents and depends on the $J^P$ of the final state particles. The collected
expressions of $\xi$ are shown as follows:
$$
\xi=\left\{
\begin{array}{cccc}+\frac{2M_X^2}{(m_b-m_c)(m_c+m_q)},&\mathrm{for}~~DX(0^-)&-\frac{2M_X^2}{(m_b-m_c)(m_q-m_c)},&\mathrm{for}~~DX(0^+)\\
-\frac{2M_X^2}{(m_b+m_c)(m_c+m_q)},&\mathrm{for}~~D^*X(0^-)&+\frac{2M_X^2}{(m_b+m_c)(m_q-m_c)},&\mathrm{for}~~D^*X(0^+)
\end{array}\right.
$$
and $0$ for others, where $X$ denotes a $D_q\ (q=s,d)$ meson with
its $J^P$ shown in the bracket. The current quark masses encountered
in $G(m_p,k^2)$ and $\xi$ are taken from Ref.~\cite{PDG} and then
evolved to the scale $\mu\sim m_b$ by the renormalization group
equation of the running quark masses.

The amplitudes for double charmed $B$ decays considered here can
be written as $\mathcal{M}=V_{cb}V^*_{cq}T_1+V_{ub}V^*_{uq}T_2.$
Then the direct CP asymmetry can be written as
$\mathcal{A}_{cp}\equiv \frac{\Gamma(\bar{B}\rightarrow
\bar{f})-\Gamma(B\rightarrow f)}{\Gamma(\bar{B}\rightarrow
\bar{f})+\Gamma(B\rightarrow f)}
=D_1\frac{\sin\gamma}{1+D_2\cos\gamma}$ where $\bar{B} (B)$
denotes a meson with a $b$ quark ($\bar{b}$ anti-quark), $\bar{f}$
is the CP conjugated state of $f$. The weak phase
$\gamma\equiv\arg(-\frac{V^*_{ub}V_{ud}}{V^*_{cb}V_{cd}})\simeq\arg(\frac{V^*_{ub}V_{us}}{V^*_{cb}V_{cs}})$.
$D_1\equiv\frac{\epsilon_i2\sin\delta}{|G_1/G_2|+|G_2/G_1|}$ and
$D_2\equiv\frac{\epsilon_i2\cos\delta}{|G_1/G_2|+|G_2/G_1|}$ where
$G_1=V_{cb}V^*_{cq}T_1$, $G_2=V_{ub}V^*_{uq}T_2$ and the strong
phase $\delta=\arg(T_2)-\arg(T_1)$. $\epsilon_1=+1$ for $q=s$, and
$\epsilon_2=-1$ for $q=d$, respectively. For numerical
calculations, we need the values of following input parameters:
\begin{equation}\begin{array}{ccccc}
C_1=1.0849,&C_2=-0.1902,&C_3=0.0148,&C_4=-0.0362,&C_5=0.0088,\\
C_6=-0.0422,&\frac{C_7}{\alpha_{e}}=-0.0007,&\frac{C_8}{\alpha_{e}}=0.0565,&\frac{C_9}{\alpha_{e}}=-1.3039,&\frac{C_{10}}{\alpha_{e}}=0.2700.
\end{array}
\end{equation}
The Wilson coefficients at the scale $\mu\sim m_b$ and coupling
constants are quoted from Ref.~\cite{Sun}, and the others like CKM
matrix elements and life times of mesons are taken from PDG
~\cite{PDG}. The decay constants used are shown in
Table~\ref{tab4-1}, $\alpha_s(m_b)=0.216$.

The estimated branching ratios of $B\rightarrow D^{(*)}l\nu_l$
decays are listed in Table~\ref{tab4-2}. The results are generally
consistent with each other. Our uncertainties are obtained by
varying the input parameters by $\pm10\%$. The decay rates for
$B\rightarrow D^{(*)}X$ are shown in Table \ref{tab4-3}. The
results from ``DBGN", ``NS" and ``CGW" are estimated within FA or
the generalized factorization approach (GFA). The $B\rightarrow
D^{(*)}X$ decay amplitudes discussed here evaluated within FA, GFA
and QCDF have the same structure. For color-favored dominated
processes, $a_1$ won't vary too much from method to method.
Actually for the decays discussed here $a_1\sim1-1.1$. So the
differences among these results could reflect the differences on
$B\rightarrow D^{(*)}$ form factors. From Table \ref{tab4-3}, we
can see that our results are roughly consistent with those from
other methods and within the error bars of experimental data.

The direct CP asymmetries are shown in Table~\ref{tab4-4}. We can
see that the CPA estimated with FA are generally within the
experiment errors, but are quite different from the pQCD
predictions. In pQCD, the annihilation diagram contributes the
leading strong phase~\cite{LiHN}, whereas in FA this diagram is
totally ignored. Recent years, the Belle Collaboration and the
BaBar Collaboration measured the CPA of $\bar{B}^0\rightarrow
D^+D^-$ mode~\cite{Belle2,BaBar}: $
\mathcal{A}_{CP}=0.07\pm0.23\pm0.03$ (BaBar),
$\mathcal{A}_{CP}=0.91\pm0.23\pm0.06$ (Belle). If the measurement
of Belle is justified, it cannot be explained by either the FA or
pQCD.

We turn to discuss the polarization fractions of $B\rightarrow
VV(A)$ decays, which are defined as
$R_i=\frac{|\mathcal{M}_i|^2}{|\mathcal{M}_L|^2+|\mathcal{M}_{\parallel}|^2+|\mathcal{M}_{\perp}|^2}$,
where $\mathcal{M}_i$, $\mathcal{M}_\parallel$ and
$\mathcal{M}_\perp$ are the longitudinal, transverse parallel and
transverse perpendicular part of the amplitude respectively. The
results are listed in Table~\ref{tab4-5}. Our results agree well
with the experimental data except the $R_\perp$ in
$\bar{B}^0\rightarrow D^{*+}D^{*-}$ mode. We note that the pQCD
prediction on $R_\perp\sim0.6$~\cite{PQCD2} in $\bar{B}^0\rightarrow
D^{*+}D^{*-}$ are not consistent with the experimental data either.
So a large $R_\perp$ cannot be explained by the non-factorizable
effects, at least in the framework of pQCD. According to our
results, for $B\rightarrow DL$ decays, the relation $R_L\sim 0.8\gg
R_\parallel\gg R_\perp$ holds; for $B\rightarrow DD$ decays $R_L\sim
R_\parallel\sim0.5\gg R_\perp$ holds.

In summary, within the framework of FA, we use the Salpeter method
to study the decays of $B\rightarrow D^{(*)}l\nu_{l}$ and
$B\rightarrow D^{(*)}X$. The direct CP asymmetries do not
contradict the measurements. For the polarization fractions, our
predictions on $R_L$ agree well with the experimental
measurements, but on $R_\perp$ in $\bar{B}^0\rightarrow
D^{*+}D^{*-}$ mode, the experimental data are larger than our
results.

\begin{table} \tiny\caption{Decay constants used in our calculation in unit of MeV.}
\begin{center}\begin{tabular}{|c|c|c|c|c|c|c|}
\hline\hline$f_{\pi}$&$f_{k}$&$f_{D}$&$f_{D_s}$&$f_{\rho}$&$f_{k^*}$
&$f_{a_1}$\\\hline 130 \cite{PDG}&156 \cite{PDG}&$207^{+9}_{-9}$
\cite{PDG}&$258^{+6}_{-6}$ \cite{PDG}&$205^{+9}_{-9}$
\cite{Ball}&$217^{+5}_{-5}$ \cite{Ball}&$229^{+10}_{-10}$
\cite{NS}\\\hline\hline
$f_{D^*}$&$f_{D_s^*}$&$f_{D^*_0}$&$f_{D_{s0}^*}$&$f_{D_{s1}(2460)}$&$f_{D_{s1}(2536)}$
&\\\hline $245^{+20}_{-20}$ \cite{Bec}&$272^{+16}_{-16}$\cite{Bec}
&$137^{+4}_{-5}~\cite{Fu}$&$109^{+4}_{-5}~\cite{Fu}$&$227^{+22}_{-19}~\cite{Fu}$&$77.3^{+12.4}_{-9.8}~\cite{Fu}$&\\\hline
\end{tabular}\end{center}\label{tab4-1}
\end{table}

\begin{table} \tiny\caption{Branching ratios of semi-leptonic $B\rightarrow D^{(*)}l\nu_l$
decays.}
\begin{center}\begin{tabular}{|c|c|c|c|c|}
\hline\hline Processes&This work&CQM~\cite{CQM3}&QCDSR~\cite{Azizi}&Experiment~\cite{PDG} \\
\hline $B^0\rightarrow D^{-}l^+\nu_l~(l=e,\mu)$&$(1.3-2.0)~\%$&$(2.2-3.0)\%$&$(1.5-2.4)\%$&$(2.17\pm0.12)~\%$\\
\hline $B^0\rightarrow
D^{-}\tau^+\nu_{\tau}$&$(4.0-6.1)\times10^{-3}$&&&$(1.1\pm0.4)\%$\\\hline
\hline $B^0\rightarrow
D^{*-}l^+\nu_l~(l=e,\mu)$&$(4.7-6.0)~\%$&$(5.9-7.6)\%$&$(4.36-8.94)\%$&$(5.01\pm0.12)\%$\\
\hline $B^0\rightarrow
D^{*-}\tau^+\nu_{\tau}$&$(1.1-1.4)~\%$&&&$(1.5\pm0.5)\%$\\\hline
\hline $B^+\rightarrow \bar{D}^{0}l^+\nu_l~(l=e,\mu)$&$(1.4-2.2)~\%$&$(2.2-3.0)\%$&$(1.3-2.2)\%$&$(2.23\pm0.11)~\%$\\
\hline $B^+\rightarrow
\bar{D}^{0}\tau^+\nu_{\tau}$&$(4.3-6.6)\times10^{-3}$&&&$(7\pm4)\times10^{-3}$\\\hline
\hline $B^+\rightarrow
\bar{D}^{*0}l^+\nu_l~(l=e,\mu)$&$(5.0-6.5)~\%$&$(5.9-7.6)\%$&$(4.57-9.12)\%$&$(5.68\pm0.19)\%$\\
\hline $B^+\rightarrow
\bar{D}^{*0}\tau^+\nu_{\tau}$&$(1.2-1.5)~\%$&&&$(2.0\pm0.5)\%$\\\hline
\end{tabular}\end{center}\label{tab4-2}
\end{table}

\begin{table} \tiny\caption{The (averaged) branching ratios of $\bar{B}^0\rightarrow
D^{(*)+}X^-$ and $B^-\rightarrow D^{(*)0}D_{(s)}^-$ in unit of
$10^{-3}$.}
\begin{center}\begin{tabular}{|c|c|c|c|c|c|c|c|} \hline\hline
channels&This work&DBGN~\cite{DBGN} &NS~\cite{NS}&CGW~\cite{CGW}
&QCDF~\cite{QCDF1}&pQCD~\cite{PQCD3}& Ex~\cite{PDG}\\\hline
$D^+\pi^-$&$3.2^{+0.4}_{-0.8}$&2.5&3.0&&3.27&$2.69^{+0.75}_{-0.66}$&$2.68\pm0.13$\\\hline
$D^+\rho^-$&$7.5^{+1.7}_{-2.3}$&6.6&7.5&&7.64&$6.70^{+1.88}_{-1.65}$&$7.6\pm1.3$\\\hline
$D^+k^-$&$0.24^{+0.03}_{-0.05}$&0.20&0.2&&0.25&$0.227^{+0.064}_{-0.056}$&$0.20\pm0.06$\\\hline
$D^+k^{*-}$&$0.44^{+0.08}_{-0.12}$&0.33&0.4&&0.39&$0.383^{+0.107}_{-0.094}$&$0.45\pm0.07$\\\hline
$D^+a_1^{-}$&$8.7^{+1.9}_{-2.8}$&&8.1&&7.76&&$6.0\pm3.3$\\\hline
\hline
$D^{*+}\pi^-$&$3.7^{+0.3}_{-0.9}$&2.6&2.8&&3.05&$2.61^{+0.73}_{-0.64}$&$2.76\pm0.13$\\\hline
$D^{*+}\rho^-$&$10.3^{+1.7}_{-3.0}$&8.7&8.4&&7.59&$7.54^{+2.11}_{-1.85}$&$6.8\pm0.9$\\\hline
$D^{*+}k^-$&$0.28^{+0.02}_{-0.07}$&0.20&0.2&&0.22&$0.221^{+0.062}_{-0.054}$&$0.214\pm0.016$\\\hline
$D^{*+}k^{*-}$&$0.64^{+0.07}_{-0.17}$&0.45&0.5&&0.40&$0.463^{+0.130}_{-0.114}$&$0.33\pm0.06$\\\hline
$D^{*+}a_1^{-}$&$15^{+2}_{-4}$&&11.6&&8.53&&$13.0\pm2.7$\\\hline
\hline
$D^+D^-$&$0.35^{+0.07}_{-0.11}$&0.42&0.3&0.31&&$0.37^{+0.29}_{-0.18}$&$0.211\pm0.031$\\\hline
$D^+D^{*-}$&$0.38^{+0.12}_{-0.15}$&0.28&0.3&0.42&&$0.32^{+0.24}_{-0.15}$&$0.61\pm0.15$\\\hline
$D^{+}D_s^-$&$9.3^{+1.2}_{-2.6}$&10&9.6&8.25&&$8.9^{+6.8}_{-4.3}$&$7.2\pm0.8$\\\hline
$D^{+}D_s^{*-}$&$7.8^{+2.1}_{-2.9}$&6.8&8.0&10.80&&$8.3^{+6.1}_{-3.9}$&$7.4\pm1.6$\\\hline
$D^+D^{*-}_0$&$0.13^{+0.02}_{-0.04}$&&&&&&\\\hline
$D^+D^{*-}_{s0}$&$1.4^{+0.2}_{-0.5}$&&&&&&\\\hline
$D^{+}D_{s1}^-(2460)$&$4.0^{+1.6}_{-1.7}$&&&&&&\\\hline
$D^{+}D_{s1}^{-}(2536)$&$0.43^{+0.23}_{-0.21}$&&&&&&\\\hline\hline
$D^{*+}D^-$&$0.36^{+0.06}_{-0.12}$&0.31&0.2&0.29&&$0.46^{+0.35}_{-0.21}$&\\\hline
$D^{*+}D^{*-}$&$1.16^{+0.22}_{-0.39}$&1.0&0.8&0.91&&$0.63^{+0.48}_{-0.30}$&$0.82\pm0.09$\\\hline
$D^{*+}D_s^-$&$9.5^{+1.3}_{-2.9}$&7.0&5.8&7.67&&$11.2^{+8.6}_{-5.3}$&$8.0\pm1.1$\\\hline
$D^{*+}D_s^{*-}$&$26^{+3}_{-8}$&26&23.2&25.51&&$16.8^{+13.0}_{-8.8}$&$17.7\pm1.4$\\\hline
$D^{*+}D_{0}^{*-}$&$0.12^{+0.01}_{-0.04}$&&&&&&\\\hline
$D^{*+}D_{s0}^{*-}$&$1.4^{+0.2}_{-0.5}$&&&&&&\\\hline
$D^{*+}D_{s1}^-(2460)$&$19^{+4}_{-7}$&&&&&&\\\hline
$D^{*+}D_{s1}^{-}(2536)$&$2.2^{+0.7}_{-0.9}$&&&&&&\\\hline\hline
$D^0D^{-}$&$0.38^{+0.07}_{-0.12}$&0.42&0.4&0.33&&$0.39^{+0.29}_{-0.19}$&$0.38\pm0.04$\\\hline
$D^0D^{*-}$&$0.41^{+0.14}_{-0.16}$&0.28&0.3&0.45&&$0.36^{+0.26}_{-0.17}$&$0.39\pm0.05$\\\hline
$D^0D_s^{-}$&$10.0^{+1.3}_{-2.8}$&10&10.3&8.94&&$9.5^{+6.9}_{-4.6}$&$10.0\pm1.7$\\\hline
$D^0D_s^{*-}$&$8.4^{+2.3}_{-3.1}$&6.8&8.5&11.73&&$8.9^{+6.4}_{-4.2}$&$7.6\pm1.6$\\\hline
$D^0D_0^{*-}$&$0.14^{+0.02}_{-0.04}$&&&&&&\\\hline
$D^0D_{s0}^{*-}$&$1.5^{+0.2}_{-0.5}$&&&&&&\\\hline
$D^0D_{s1}^{-}(2460)$&$4.4^{+1.6}_{-1.9}$&&&&&&\\\hline
$D^0D_{s1}^{-}(2536)$&$0.46^{+0.26}_{-0.22}$&&&&&&\\\hline\hline
$D^{*0}D^{-}$&$0.39^{+0.07}_{-0.13}$&0.31&0.2&0.31&&$0.48^{+0.34}_{-0.23}$&$0.63\pm0.17$\\\hline
$D^{*0}D^{*-}$&$1.25^{+0.23}_{-0.43}$&1.0&0.9&0.98&&$0.68^{+0.50}_{-0.32}$&$0.81\pm0.17$\\\hline
$D^{*0}D_s^{-}$&$10.3^{+1.3}_{-3.2}$&7.0&6.2&8.34&&$11.9^{+9.4}_{-5.6}$&$8.2\pm1.7$\\\hline
$D^{*0}D_s^{*-}$&$28^{+4}_{-9}$&26&24.7&27.69&&$18.1^{+13.9}_{-9.5}$&$17.1\pm2.4$\\\hline
$D^{*0}D_0^{*-}$&$0.13^{+0.02}_{-0.05}$&&&&&&\\\hline
$D^{*0}D_{s0}^{*-}$&$1.5^{+0.2}_{-0.5}$&&&&&&\\\hline
$D^{*0}D_{s1}^{-}(2460)$&$21^{+3}_{-8}$&&&&&&\\\hline
$D^{*0}D_{s1}^{-}(2536)$&$2.4^{+0.8}_{-1.0}$&&&&&&\\\hline
\end{tabular}\end{center} \label{tab4-3}
\end{table}

\begin{table}[h] \tiny\caption{The direct CP asymmetries and $D_1$, $D_2$ in double charmed non-leptonic decays of
$\bar{B}^0$ and $B^-$. The weak phase is taken as
$\gamma=68.8^{\circ}$.}
{\begin{center}\begin{tabular}{|c|c|c|c|c|c|} \hline Final
States&$D_1$&$D_2$&$\mathcal{A}_{CP}$&$\mathcal{A}_{CP}$(PQCD)~\cite{PQCD3}&$\mathcal{A}_{CP}$(Ex)~\cite{PDG}
\\\hline
$D^+D^{-}$&-0.053&0.10&-0.048&$(0.5^{+0.1+0.5}_{-0.2-0.4})\times10^{-2}$&$0.5\pm0.4$\\\hline
$D^+D^{*-}$&-0.015&0.026&-0.014&$\sim10^{-4}$&$0.02\pm0.04$\\\hline
$D^{+}D_s^{-}$&0.0032&-0.0097&0.0030&&\\\hline
$D^{+}D_s^{*-}$&0.00081&-0.0014&0.00076&&\\\hline
$D^+D^{*-}_0$&-0.083&0.17&-0.073&&\\\hline
$D^+D^{*-}_{s0}$&0.0050&-0.010&0.0047&&\\\hline
$D^{+}D_{s1}^-(2460)$&0.00082&-0.0016&0.00076&&\\\hline
$D^{+}D_{s1}^{-}(2536)$&0.00082&-0.0017&0.00076&&\\\hline\hline
$D^{*+}D^{-}$&0.0046&-0.016&0.0043&$(-0.6^{+0.0+0.1}_{-0.1-0.2})\times10^{-2}$&\\\hline
$D^{*+}D^{*-}$&-0.015&0.026&-0.014&$\sim-10^{-4}$&$0.01\pm0.09$\\\hline
$D^{*+}D_s^{-}$&-0.00027&0.0013&-0.00025&&\\\hline
$D^{*+}D_s^{*-}$&0.00081&-0.0014&0.00076&&\\\hline
$D^{*+}D_{0}^{*-}$&0.014&-0.043&0.013&&\\\hline
$D^{*+}D_{s0}^{*-}$&-0.00086&0.0027&-0.00080&&\\\hline
$D^{*+}D_{s1}^-(2460)$&0.00082&-0.0016&0.00076&&\\\hline
$D^{*+}D_{s1}^{-}(2536)$&0.00082&-0.0017&0.00076&&\\\hline\hline
$D^0D^-$&-0.053&0.10&-0.048&$(0.6^{+0.4+0.4}_{-0.0-0.1})\times10^{-2}$&$-0.03\pm0.07$\\\hline
$D^0D^{*-}$&-0.015&0.026&-0.014&$(0.1^{+0.4+0.1}_{-0.1-0.1})\times10^{-2}$&$-0.06\pm0.13$\\\hline
$D^{0}D_s^{-}$&0.0032&-0.0097&0.0030&$\sim-10^{-5}$&\\\hline
$D^{0}D_s^{*-}$&0.00081&-0.0014&0.00076&$\sim-10^{-5}$&\\\hline
$D^0D_0^{*-}$&-0.083&0.17&-0.073&&\\\hline
$D^0D_{s0}^{*-}$&0.0050&-0.010&0.0047&&\\\hline
$D^0D_{s1}^{-}(2460)$&0.00082&-0.0016&0.00076&&\\\hline
$D^0D_{s1}^{-}(2536)$&0.00082&-0.0017&0.00076&&\\\hline\hline
$D^{*0}D^{-}$&0.0046&-0.016&0.0043&$(-0.5_{-0.2-0.3}^{+0.1+0.0})\times10^{-2}$&$0.13\pm0.18$\\\hline
$D^{*0}D^{*-}$&-0.015&0.026&-0.014&$(0.2_{-0.1-0.1}^{+0.0+0.0})\times10^{-2}$&$-0.15\pm0.11$\\\hline
$D^{*0}D_s^{-}$&-0.00027&0.0013&-0.00025&$\sim-10^{-5}$&\\\hline
$D^{*0}D_s^{*-}$&0.00081&-0.0014&0.00076&$\sim-10^{-5}$&\\\hline
$D^{*0}D_0^{*-}$&0.014&-0.043&0.013&&\\\hline
$D^{*0}D_{s0}^{*-}$&-0.00086&0.0027&-0.00080&&\\\hline
$D^{*0}D_{s1}^{-}(2460)$&0.00082&-0.0016&0.00076&&\\\hline
$D^{*0}D_{s1}^{-}(2536)$&0.00082&-0.0017&0.00076&&\\\hline
\end{tabular}\end{center} \label{tab4-4}}
\end{table}

\begin{table}[h] \tiny\caption{Polarization fractions of $B\rightarrow VV$ or
$VA$ decays. $R_L$, $R_{\parallel}$ and $R_{\perp}$ are
longitudinal, transverse parallel and transverse perpendicular
polarization fractions, respectively. }
{\begin{center}\begin{tabular}{|c|c|c|c|c|c|c|} \hline Final
States&$R_L$&$R_{\parallel}$&$R_{\perp}$&$R_L$ in
pQCD~\cite{PQCD3}&$R_L$ in experiment~\cite{PDG}&$R_\perp$ in
experiment~\cite{PDG}
\\\hline
$D^{*+}\rho^-$&$0.878^{+0.007}_{-0.008}$&$0.101^{+0.009}_{-0.008}$&$0.021^{+0.002}_{-0.001}$&0.85&$0.885\pm0.016\pm0.012$&\\\hline
$D^{*+}K^{*-}$&$0.845^{+0.008}_{-0.009}$&$0.128^{+0.012}_{-0.010}$&$0.026^{+0.002}_{-0.001}$&0.81&&\\\hline
$D^{*+}a_1^-$&$0.744^{+0.011}_{-0.013}$&$0.215^{+0.016}_{-0.015}$&$0.041^{+0.004}_{-0.003}$&&&\\\hline\hline
$D^{*+}D^{*-}$&$0.532^{+0.011}_{-0.013}$&$0.409^{+0.019}_{-0.019}$&$0.059^{+0.009}_{-0.007}$&$0.54\pm0.14$&$0.57\pm0.08\pm0.02$&$0.150\pm0.025$\\\hline
$D^{*+}D_s^{*-}$&$0.509^{+0.010}_{-0.012}$&$0.432^{+0.019}_{-0.019}$&$0.059^{+0.009}_{-0.007}$&$0.52^{+0.14}_{-0.13}$&$0.52\pm0.05$&\\\hline
$D^{*+}D_{s1}^{-}(2460)$&$0.441^{+0.008}_{-0.008}$&$0.506^{+0.017}_{-0.017}$&$0.053^{+0.010}_{-0.008}$&&&\\\hline
$D^{*+}D_{s1}^{-}(2536)$&$0.428^{+0.007}_{-0.008}$&$0.521^{+0.016}_{-0.016}$&$0.050^{+0.010}_{-0.008}$&&&\\\hline
$D^{*0}D^{*-}$&$0.533^{+0.010}_{-0.013}$&$0.408^{+0.020}_{-0.019}$&$0.059^{+0.009}_{-0.007}$&$0.55\pm0.13$&&\\\hline
$D^{*0}D_s^{*-}$&$0.510^{+0.010}_{-0.012}$&$0.431^{+0.019}_{-0.019}$&$0.059^{+0.009}_{-0.007}$&$0.52^{+0.14}_{-0.12}$&&\\\hline
$D^{*0}D_{s1}^{-}(2460)$&$0.442^{+0.007}_{-0.009}$&$0.505^{+0.017}_{-0.017}$&$0.053^{+0.010}_{-0.008}$&&&\\\hline
$D^{*0}D_{s1}^{-}(2536)$&$0.429^{+0.007}_{-0.008}$&$0.521^{+0.016}_{-0.017}$&$0.050^{+0.010}_{-0.007}$&&&\\\hline
\end{tabular}\end{center} \label{tab4-5}}
\end{table}

\section*{Acknowledgments}

The work of G.W. was supported by NSFC under Grant No.~10875032,
No.~11175051, and No.~10911140267. The work of X.C. was supported
by the Foundation of Harbin Institute of Technology (Weihai)
No.~IMJQ~10000076.

\end{document}